\documentclass[12pt]{article}

\usepackage{graphicx}

\usepackage{amssymb}

 \usepackage{amsmath}
\usepackage{siunitx}

\title{Covalent Adsorption of functional groups on [N]-carbophenes}

\author{
        Chad E. Junkermeier \\
                Department of Science, Technology, Engineering, and Mathematics\\ University of Hawai`i Maui College\\
                Kahului HI 96732, {USA}
            \and
        George Psofogiannakis\\
        Department of Chemical and Biological Engineering\\ University of Ottawa\\ Ottawa, Ontario, {Canada}
            \and
            Ricardo Paupitz\\
            Departamento de F\'{\i}sica, IGCE\\ Universidade Estadual Paulista\\ UNESP, 13506-900\\ Rio Claro, SP, Brazil
}

\begin{document}

\maketitle

%%%%%%%%%%%%%%%%%%%%%%%%%%%%%%%%%%%%%%%%%%%%%%%%%%%%%%%%%%%%%%%%%%%%%%%%%%%%%%
%%%%%%%%%%%%%%%%%%%%%%%%%%%%%%%%%%%%%%%%%%%%%%%%%%%%%%%%%%%%%%%%%%%%%%%%%%%%%%
%%%%%%%%%%%%%%%%%%%%%%%%%%%%%%%%%%%%%%%%%%%%%%%%%%%%%%%%%%%%%%%%%%%%%%%%%%%%%%
%%%%%%%%%%%%%%%%%%%%%%%%%%%%%%%%%%%%%%%%%%%%%%%%%%%%%%%%%%%%%%%%%%%%%%%%%%%%%%
%%%%%%%%%%%%%%%%%%%%%%%%%%%%%%%%%%%%%%%%%%%%%%%%%%%%%%%%%%%%%%%%%%%%%%%%%%%%%%
%%%%%%%%%%%%%%%%%%%%%%%%%%%%%%%%%%%%%%%%%%%%%%%%%%%%%%%%%%%%%%%%%%%%%%%%%%%%%%
\begin{abstract}
%%%%%%%%%%%%%%%%%%%%%%%%%%%%%%%%%%%%%%%%%%%%%%%%%%%%%%%%%%%%%%%%%%%%%%%%%%%%%%
%%%%%%%%%%%%%%%%%%%%%%%%%%%%%%%%%%%%%%%%%%%%%%%%%%%%%%%%%%%%%%%%%%%%%%%%%%%%%%
%%%%%%%%%%%%%%%%%%%%%%%%%%%%%%%%%%%%%%%%%%%%%%%%%%%%%%%%%%%%%%%%%%%%%%%%%%%%%%
%%%%%%%%%%%%%%%%%%%%%%%%%%%%%%%%%%%%%%%%%%%%%%%%%%%%%%%%%%%%%%%%%%%%%%%%%%%%%%
%%%%%%%%%%%%%%%%%%%%%%%%%%%%%%%%%%%%%%%%%%%%%%%%%%%%%%%%%%%%%%%%%%%%%%%%%%%%%%
%%%%%%%%%%%%%%%%%%%%%%%%%%%%%%%%%%%%%%%%%%%%%%%%%%%%%%%%%%%%%%%%%%%%%%%%%%%%%%

Starting from the planar molecule 1,3,5-trihydroxybenzene, Du \textit{et al.} reported synthesizing one of a couple of possible 2D materials: graphenylene or 3-carbophene.\cite{du1740796}  3-carbophene is a member of a novel class of two-dimensional covalent organic framework, [N]-carbophenes (carbophenes).  Using a high throughput method, we computed the formation energies and conduction properties of 3-, 4-, 5-, and 6-carbophenes with hydroxyl (OH), carbonyl (CO), nitro (NO$_2$), amine (NH$_2$), carboxyl (COOH) functional groups replacing hydrogen terminating agents. Five hundred and nine structures with randomly picked motifs, with functionalizations from a single functional group per cell to fully functionalized were studied.  Our results demonstrate a negatively sloped linear relationship between the degree of functionalization and formation energy when the type of functional group and type of carbophene are held constant.  The decrease in formation energy with functionalization makes Du's synthesis of functionalized 3-carbophene more creditable.  
The type of carbophene, type of functional group, and the degree of functionalization all play a role in the band structure of the materials.  For example, CO functional groups may lead to a mid-gap state pinned to the Fermi level, whereas the other functional groups studied keep the semiconducting nature of pristine carbophene.  Thus, because carbonyl functional groups are often present in defected carbon systems, care should be taken to limit the amount of oxygen in carbophene devices where the band gap is important.
Thus, this work strengthens the hypothesis of Junkermeier \textit{et al.}'s hypothesis that Du \textit{et al.} synthesized 3-carbophene and not graphenylene.\cite{junkermeier2019simplecarbophene}

%%%%%%%%%%%%%%%%%%%%%%%%%%%%%%%%%%%%%%%%%%%%%%%%%%%%%%%%%%%%%%%%%%%%%%%%%%%%%%
%%%%%%%%%%%%%%%%%%%%%%%%%%%%%%%%%%%%%%%%%%%%%%%%%%%%%%%%%%%%%%%%%%%%%%%%%%%%%%
%%%%%%%%%%%%%%%%%%%%%%%%%%%%%%%%%%%%%%%%%%%%%%%%%%%%%%%%%%%%%%%%%%%%%%%%%%%%%%
%%%%%%%%%%%%%%%%%%%%%%%%%%%%%%%%%%%%%%%%%%%%%%%%%%%%%%%%%%%%%%%%%%%%%%%%%%%%%%
%%%%%%%%%%%%%%%%%%%%%%%%%%%%%%%%%%%%%%%%%%%%%%%%%%%%%%%%%%%%%%%%%%%%%%%%%%%%%%
%%%%%%%%%%%%%%%%%%%%%%%%%%%%%%%%%%%%%%%%%%%%%%%%%%%%%%%%%%%%%%%%%%%%%%%%%%%%%%
\end{abstract}
%%%%%%%%%%%%%%%%%%%%%%%%%%%%%%%%%%%%%%%%%%%%%%%%%%%%%%%%%%%%%%%%%%%%%%%%%%%%%%
%%%%%%%%%%%%%%%%%%%%%%%%%%%%%%%%%%%%%%%%%%%%%%%%%%%%%%%%%%%%%%%%%%%%%%%%%%%%%%
%%%%%%%%%%%%%%%%%%%%%%%%%%%%%%%%%%%%%%%%%%%%%%%%%%%%%%%%%%%%%%%%%%%%%%%%%%%%%%
%%%%%%%%%%%%%%%%%%%%%%%%%%%%%%%%%%%%%%%%%%%%%%%%%%%%%%%%%%%%%%%%%%%%%%%%%%%%%%
%%%%%%%%%%%%%%%%%%%%%%%%%%%%%%%%%%%%%%%%%%%%%%%%%%%%%%%%%%%%%%%%%%%%%%%%%%%%%%
%%%%%%%%%%%%%%%%%%%%%%%%%%%%%%%%%%%%%%%%%%%%%%%%%%%%%%%%%%%%%%%%%%%%%%%%%%%%%%

\section{Introduction}
%%%%%%%%%%%%%%%%%%%%%%%%%%%%%%%%%%%%%%%%%%%%%%%%%%%%%%%%%%%%%%%%%%%%%%%%%%%%%%
%%%%%%%%%%%%%%%%%%%%%%%%%%%%%%%%%%%%%%%%%%%%%%%%%%%%%%%%%%%%%%%%%%%%%%%%%%%%%%
%%%%%%%%%%%%%%%%%%%%%%%%%%%%%%%%%%%%%%%%%%%%%%%%%%%%%%%%%%%%%%%%%%%%%%%%%%%%%%
%%%%%%%%%%%%%%%%%%%%%%%%%%%%%%%%%%%%%%%%%%%%%%%%%%%%%%%%%%%%%%%%%%%%%%%%%%%%%%
%%%%%%%%%%%%%%%%%%%%%%%%%%%%%%%%%%%%%%%%%%%%%%%%%%%%%%%%%%%%%%%%%%%%%%%%%%%%%%
%%%%%%%%%%%%%%%%%%%%%%%%%%%%%%%%%%%%%%%%%%%%%%%%%%%%%%%%%%%%%%%%%%%%%%%%%%%%%%

The characterization of graphene has produced an ever strengthening tidal wave of research into two-dimensional (2D) materials.\cite{Bhimanapati20155b05556} 
To overcome the technical limitations for its practical use, an increasing number of modifications have been proposed, such as the application of strain, combination with substrate, oxidation, for instance.\cite{Wolf011488,Giovannetti07073103,Robinson083137,Fujita10043508,Casolo09054704,SanchezYamagishi2012076601}. Several authors have also proposed and investigated, using both experimental and {\it ab initio} methods, the use of functionalization as a tool to modulate mechanical, electronic and optical properties of graphene, graphene oxide, graphene quantum dots and many other similar systems.\cite{functional_groups-luminescence,functional_groups_optical_properties,functional_groups_first_principles,functional_groups_dft,Kamran2019234}.
In addition to graphene, numerous 2D carbon allotropes are under investigation theoretically,\cite{Zhu19945281,graphynefamily1998,Lu133677,Zhang10813401,Yang20206379} with several having been synthesized.\cite{peng2014new,LI2018248,Fan2021852} Many inorganic 2D materials have also been proposed, with several being produced in the lab: silicene, germanene, phosphorene, hexagonal boron nitride (h-BN), gallium nitride (GaN$_{x}$), gallium selenide (GaSe$_x$), GaO$_x$, tungsten disulfide (WS$_2$), tungsten disulfide(WSe$_2$), molybdenum disulfide (MoS$_2$), molybdenum diselenide (MoS$_2$), molybdenum ditelluride (MoTe$_2$), rhenium disulfide (ReS$_2$).\cite{Bhimanapati20155b05556,Balendhran2015201402041,Das20151,Rahman2017201606129,Liu20202001232}

A previous work discussed [N]-carbophenes (carbophenes), a novel class of two-dimensional covalent organic frameworks based on linear [N]-phenylenes.\cite{junkermeier2019simplecarbophene} While this class of materials could have easily been deduced as the logical extension of graphenylene, it was not addressed in literature until its recent possible synthesis.\cite{du1740796} The goal of Du \textit{et al.} was to synthesize graphenylene. Yet, they realized that their methods could have also created carbophenes. Junkermeier \textit{et al.} demonstrated that the formation energies per carbon atom of graphenylene and pristine carbophene are similar. They further showed that carbophene has an interlayer separation that is closer to the experimental value found by Du \textit{et al.} than that of graphenylene. The pristine carbophenes used in that previous theoretical work had only H terminating agents.  Thus, to get closer to the experimental material that Du's group may have produced, we now replace hydrogen terminating agents with the functional groups hydroxyl (OH), carbonyl (CO), nitro (NO$_2$), amine (NH$_2$), carboxyl (COOH).   

Using high throughput methods, we optimized 509 functionalized carbophene systems, computed their formation energies, and electronic structures.  The formation energy results demonstrate that adding a new functional group always lowers the energy.  The formation energy results build on previous work, to further suggest that Du \textit{et al.} synthesized 3-carbophene.  Partial functionalization may lead to a mid-gap state pinned to the Fermi level, but complete functionalization leads to an open band gap.

%%%%%%%%%%%%%%%%%%%%%%%%%%%%%%%%%%%%%%%%%%%%%%%%%%%%%%%%%%%%%%%%%%%%%%%%%%%%%%
%%%%%%%%%%%%%%%%%%%%%%%%%%%%%%%%%%%%%%%%%%%%%%%%%%%%%%%%%%%%%%%%%%%%%%%%%%%%%%
%%%%%%%%%%%%%%%%%%%%%%%%%%%%%%%%%%%%%%%%%%%%%%%%%%%%%%%%%%%%%%%%%%%%%%%%%%%%%%
%%%%%%%%%%%%%%%%%%%%%%%%%%%%%%%%%%%%%%%%%%%%%%%%%%%%%%%%%%%%%%%%%%%%%%%%%%%%%%
%%%%%%%%%%%%%%%%%%%%%%%%%%%%%%%%%%%%%%%%%%%%%%%%%%%%%%%%%%%%%%%%%%%%%%%%%%%%%%
%%%%%%%%%%%%%%%%%%%%%%%%%%%%%%%%%%%%%%%%%%%%%%%%%%%%%%%%%%%%%%%%%%%%%%%%%%%%%%
\section{Methodology}\label{sec:methods}
%%%%%%%%%%%%%%%%%%%%%%%%%%%%%%%%%%%%%%%%%%%%%%%%%%%%%%%%%%%%%%%%%%%%%%%%%%%%%%
%%%%%%%%%%%%%%%%%%%%%%%%%%%%%%%%%%%%%%%%%%%%%%%%%%%%%%%%%%%%%%%%%%%%%%%%%%%%%%
%%%%%%%%%%%%%%%%%%%%%%%%%%%%%%%%%%%%%%%%%%%%%%%%%%%%%%%%%%%%%%%%%%%%%%%%%%%%%%
%%%%%%%%%%%%%%%%%%%%%%%%%%%%%%%%%%%%%%%%%%%%%%%%%%%%%%%%%%%%%%%%%%%%%%%%%%%%%%
%%%%%%%%%%%%%%%%%%%%%%%%%%%%%%%%%%%%%%%%%%%%%%%%%%%%%%%%%%%%%%%%%%%%%%%%%%%%%%
%%%%%%%%%%%%%%%%%%%%%%%%%%%%%%%%%%%%%%%%%%%%%%%%%%%%%%%%%%%%%%%%%%%%%%%%%%%%%%

Using the density functional-based tight binding (DFTB) method implemented in DFTB+ (version 18.1 - precompiled executable), we performed geometry and lattice optimizations of the cell structures, electronic structure calculations, and computed the ground state energies of the carbophenes and constituent molecules.\cite{Elstner1998,aradi2007dftb,manzano2012,Spiegelman20201710252} DFTB+ is more computationally efficient than DFT because it pre-computes two-body interactions and stores them in look-up tables (the so-called Slater-Koster files) instead of integral evaluation at run-time.
We used the {\sl matsci Slater-Koster files} that are formulated to describe materials science problems accurately.\cite{manzano2012cement,lukose2010reticular} Dispersion forces were accounted for using Lennard-Jones potentials with the parameters taken from the Universal Force Field (UFF).\cite{Rappe1992UFF} A conjugate gradient algorithm was used in geometry and cell optimization, with a maximum force difference of $10^{-5}$ Ha/Bohr and an SCC maximum tolerance of $10^{-4}$ electrons as convergence criteria. Kpoints were optimized until an 8x8x1 Monkhorst-Pack kpoint grid was used during geometry optimization, and a 96x96x1 kpoint grid was used during the density of states (DOS) calculations. The out-of-plane bounding box was set to 30~\AA.

These methods correctly compute the unit cell dimensions and band dispersion of graphene.\cite{junkermeier2019bilayers} The methods also gave unit cells and band dispersion of graphenylene that are similar to higher-level DFT results.  The methods were also used to predict the bond lengths, valence bond angles, and out-of-plane dihedral angles of 3-phenylene that are within 1\% of experimental values and to make predictions on the nature of simple carbophenes that appear to explain what is found experimentally by Du \textit{et al.}\cite{junkermeier2019simplecarbophene}

%%%%%%%%%%%%%%%%%%%%%%%%%%%%%%%%%%%%%%%%%%%%%%%%%%%%%%%%%%%%%%%%%%%%%%%%%%%%%%
%%%%%%%%%%%%%%%%%%%%%%%%%%%%%%%%%%%%%%%%%%%%%%%%%%%%%%%%%%%%%%%%%%%%%%%%%%%%%%
%%%%%%%%%%%%%%%%%%%%%%%%%%%%%%%%%%%%%%%%%%%%%%%%%%%%%%%%%%%%%%%%%%%%%%%%%%%%%%
%%%%%%%%%%%%%%%%%%%%%%%%%%%%%%%%%%%%%%%%%%%%%%%%%%%%%%%%%%%%%%%%%%%%%%%%%%%%%%
%%%%%%%%%%%%%%%%%%%%%%%%%%%%%%%%%%%%%%%%%%%%%%%%%%%%%%%%%%%%%%%%%%%%%%%%%%%%%%
%%%%%%%%%%%%%%%%%%%%%%%%%%%%%%%%%%%%%%%%%%%%%%%%%%%%%%%%%%%%%%%%%%%%%%%%%%%%%%
\section{Results and Discussion}
%%%%%%%%%%%%%%%%%%%%%%%%%%%%%%%%%%%%%%%%%%%%%%%%%%%%%%%%%%%%%%%%%%%%%%%%%%%%%%
%%%%%%%%%%%%%%%%%%%%%%%%%%%%%%%%%%%%%%%%%%%%%%%%%%%%%%%%%%%%%%%%%%%%%%%%%%%%%%
%%%%%%%%%%%%%%%%%%%%%%%%%%%%%%%%%%%%%%%%%%%%%%%%%%%%%%%%%%%%%%%%%%%%%%%%%%%%%%
%%%%%%%%%%%%%%%%%%%%%%%%%%%%%%%%%%%%%%%%%%%%%%%%%%%%%%%%%%%%%%%%%%%%%%%%%%%%%%
%%%%%%%%%%%%%%%%%%%%%%%%%%%%%%%%%%%%%%%%%%%%%%%%%%%%%%%%%%%%%%%%%%%%%%%%%%%%%%
%%%%%%%%%%%%%%%%%%%%%%%%%%%%%%%%%%%%%%%%%%%%%%%%%%%%%%%%%%%%%%%%%%%%%%%%%%%%%%

\begin{figure}
\centering
\includegraphics[clip,width=5 in, keepaspectratio]{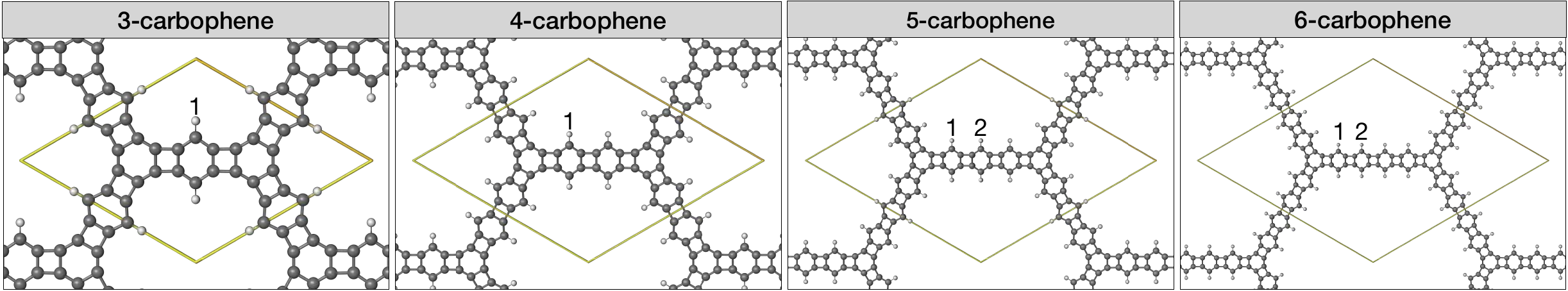}
\caption{Unit cells of pristine carbophene, (a) 3-carbophene, (b) 4-carbophene, (c) 5-carbophene, (d) 6-carbophene. Each cell labels (i.e., 1 or 2) the non-symmetric bonding sites where a single functional group replaces an H atom.}
\label{fig1:singlebondingpositions}
\end{figure}

We will first consider the effects of a substitutional replacement of a single H atom with a functional group.  When only one functional group is added, pristine 3-carbophene and 4-carbophene each have one non-symmetric bonding site under point group transformations, while 5-carbophene and 6-carbophene have two non-symmetric bonding sites each.  Figure \ref{fig1:singlebondingpositions} displays unit cells of pristine carbophenes and labels the non-symmetric binding sites for a single replacement.  The bond lengths between a bonding site carbon and the functional group were computed, see Table \ref{table:energylengths}.  The bond lengths appear to be independent of placement, indicating that bond strengths don't change between different functional sites.

%%%%%%%%%%%%%%%%%%%%%%%%%%%%%%%%%%%%%%%%%%%%%%%%%%%%%%%%%%%%%%%%%%%%%%%%%%%%%%
%%%%%%%%%%%%%%%%%%%%%%%%%%%%%%%%%%%%%%%%%%%%%%%%%%%%%%%%%%%%%%%%%%%%%%%%%%%%%%
%%%%%%%%%%%%%%%%%%%%%%%%%%%%%%%%%%%%%%%%%%%%%%%%%%%%%%%%%%%%%%%%%%%%%%%%%%%%%%
%\subsection{Formation energies}
%%%%%%%%%%%%%%%%%%%%%%%%%%%%%%%%%%%%%%%%%%%%%%%%%%%%%%%%%%%%%%%%%%%%%%%%%%%%%%
%%%%%%%%%%%%%%%%%%%%%%%%%%%%%%%%%%%%%%%%%%%%%%%%%%%%%%%%%%%%%%%%%%%%%%%%%%%%%%
%%%%%%%%%%%%%%%%%%%%%%%%%%%%%%%%%%%%%%%%%%%%%%%%%%%%%%%%%%%%%%%%%%%%%%%%%%%%%%
The formation energy ($E_F$) of each functionalized carbophene was calculated using the equation
\begin{equation}
E_F = E_{\text{carb-X}} - N_\text{C} \left(\frac{E_{\text{gr}}}{N_{\text{gr}}}\right)
      - N_\text{H} \left(\frac{E_{\text{H}_2}}{2}\right)\nonumber
      - N_\text{N} \left(\frac{E_{\text{N}_2}}{2}\right)
      - N_\text{O} \left(\frac{E_{\text{O}_2}}{2}\right),
\end{equation}
where $E_{\text{carb-X}}$ is the total energy of the carbophene-adsorbate system, $E_{\text{gr}}$ is the total energy of graphene, $N_\text{C}$ is the total number of carbon atoms in the carbophene-adsorbate system, $N_{gr}$ is the number of atoms in the pristine graphene system, $N_\text{H}$ is the total number of hydrogen atoms in the carbophene-adsorbate system, $E_{\text{H}_2}$ is the total energy of molecular hydrogen, $N_\text{N}$ is the total number of nitrogen atoms in the carbophene-adsorbate system, $E_{\text{N}_2}$ is the total energy of molecular nitrogen, $N_\text{O}$ is the total number of oxygen atoms in the carbophene-adsorbate system, and $E_{\text{O}_2}$ is the total energy of molecular oxygen. The total energy of graphene was used because it is a more relevant reference state than graphite.\cite{Song1338} Therefore, the formation energy is defined with respect to stable forms of the constituent elements, in conjunction with the usual thermodynamic definition of formation energy.

Table \ref{table:energylengths} gives the formation energies of pristine carbophenes and carbophenes with one functional group replacing an H atom.
%Pristine carbophenes have formation energies per carbon atom about 7.5\% larger than graphdiyne (0.41 eV/C atom).\cite{Li20103256}  
In the pristine cells, the change in formation energy from one size (N) of carbophene to the next is about 10 eV, indicating the smaller carbophenes are more stable.   Much of the energy that must be put into each system is due to the antiaromatic nature of 4-member rings. Thus, the rearrangement of carbon into carbophenes is understandably energy-intensive.  When a functional group replaces a hydrogen atom the formation energy of the resultant material decreases.  The type of functional group is the primary driver in the change in formation energy, but differences also occur based on bonding site.

\begin{table}[!htb]
    \centering
\caption{Formation energies (FE) and bond lengths (BL) of unit cell carbophenes with one functional group attached.}
\begin{footnotesize}    
\begin{tabular}{ccccc}
    \hline
group	&	rings	&	Site	&	FE	&	BL	\\  \hline
pristine & 3 &   & 16.09 &  1.103 \\
pristine & 4 &   & 26.47 &  1.103 \\
pristine & 5 &   & 36.95 &  1.103 \\
pristine & 6 &   & 47.45 &  1.103 \\
CO	&	3	&	1	&	15.09	&	1.366	\\
CO	&	4	&	1	&	25.59	&	1.366	\\
CO	&	5	&	1	&	36.07	&	1.366	\\
CO	&	5	&	2	&	36.17	&	1.364	\\
CO	&	6	&	1	&	46.56	&	1.365	\\
CO	&	6	&	2	&	46.67   &	1.365	\\
COOH	&	3	&	1	&	11.73	&	1.511	\\
COOH	&	4	&	1	&	22.13	&	1.509	\\
COOH	&	5	&	1	&	32.60	&	1.511	\\
COOH	&	5	&	2	&	32.62	&	1.510	\\
COOH	&	6	&	1	&	43.10	&	1.511	\\
COOH	&	6	&	2	&	43.12   &	1.510	\\
NH$_2$	&	3	&	1	&	13.95	&	1.373	\\
NH$_2$	&	4	&	1	&	24.34	&	1.372	\\
NH$_2$	&	5	&	1	&	35.60	&	1.366	\\
NH$_2$	&	5	&	2	&	34.83	&	1.373	\\
NH$_2$	&	6	&	1	&	45.31	&	1.373	\\
NH$_2$	&	6	&	2	&	45.33	&	1.372	\\
NO$_2$	&	3	&	1	&	8.19	&	1.375	\\
NO$_2$	&	4	&	1	&	18.59	&	1.375	\\
NO$_2$	&	5	&	1	&	29.07	&	1.374	\\
NO$_2$	&	5	&	2	&	29.09	&	1.374	\\
NO$_2$	&	6	&	1	&	39.56	&	1.374	\\
NO$_2$	&	6	&	2	&	39.59	&	1.374	\\
OH	&	3	&	1	&	13.62	&	1.421	\\
OH	&	4	&	1	&	26.00	&	1.380	\\
OH	&	5	&	1	&	34.51	&	1.423	\\
OH	&	5	&	2	&	34.51	&	1.422	\\
OH	&	6	&	1	&	45.01	&   1.424	\\
OH	&	6	&	2	&	46.98	&	1.381	\\  \hline
    \end{tabular}
\end{footnotesize}
    \label{table:energylengths}
\end{table}

It is interesting to note that when the functional groups are in the first bonding positions (for rings=\{5,6\}), the formation energies are lower (i.e., more stable) than when the functional groups are in the second bonding positions.  The lower energies may be due to interactions between the functional groups and the hydrogen atom at the neighboring site on the adjacent side of the hexagon.

We performed a study of the effect of various functionalization motifs (site permutations and degree of functionalization) on the formation energies of carbophenes.  The number of binding site motifs is dependent on the size, N, of carbophene. Therefore, to determine all possible functionalization motifs, we numbered each bonding site and then calculated all binding sites' possible subsets.
Several subsets with the same number of sites may be equivalent under point group transformations, in which case only one subset is kept.
Thus, 3-carbophene has six functionalizable sites, which gives 64 subset motifs in the list, 14 of which cannot be produced via reflection or rotation of a previous motif.  4-carbophene has 12 functionalizable sites, which leads to 837 motifs, and so on for 5-carbophene and 6-carbophene.  Of the remaining motifs, we selected the (already discussed) single functional motifs, the fully functionalized motifs (discussed in greater detail below), and a random selection of the other motifs.  To collect proportionally similar number of results, the probability of selection was based on the size, N, of the carbophene, with 3-carbophene motifs having a probability of 0.1, 4-carbophene motifs having a probability of 0.01, 5-carbophene a probability of 0.0001, and p=0.000002 for 6-carbophene.  The random selection of motifs was run for each type of functional group.  Thus, each functional had between 60 and 120 selected motifs.  For each of the motifs, the cell structure was optimized, and the formation energy and band gap was calculated.

\begin{figure}
\centering
\includegraphics[clip,width=5 in, keepaspectratio]{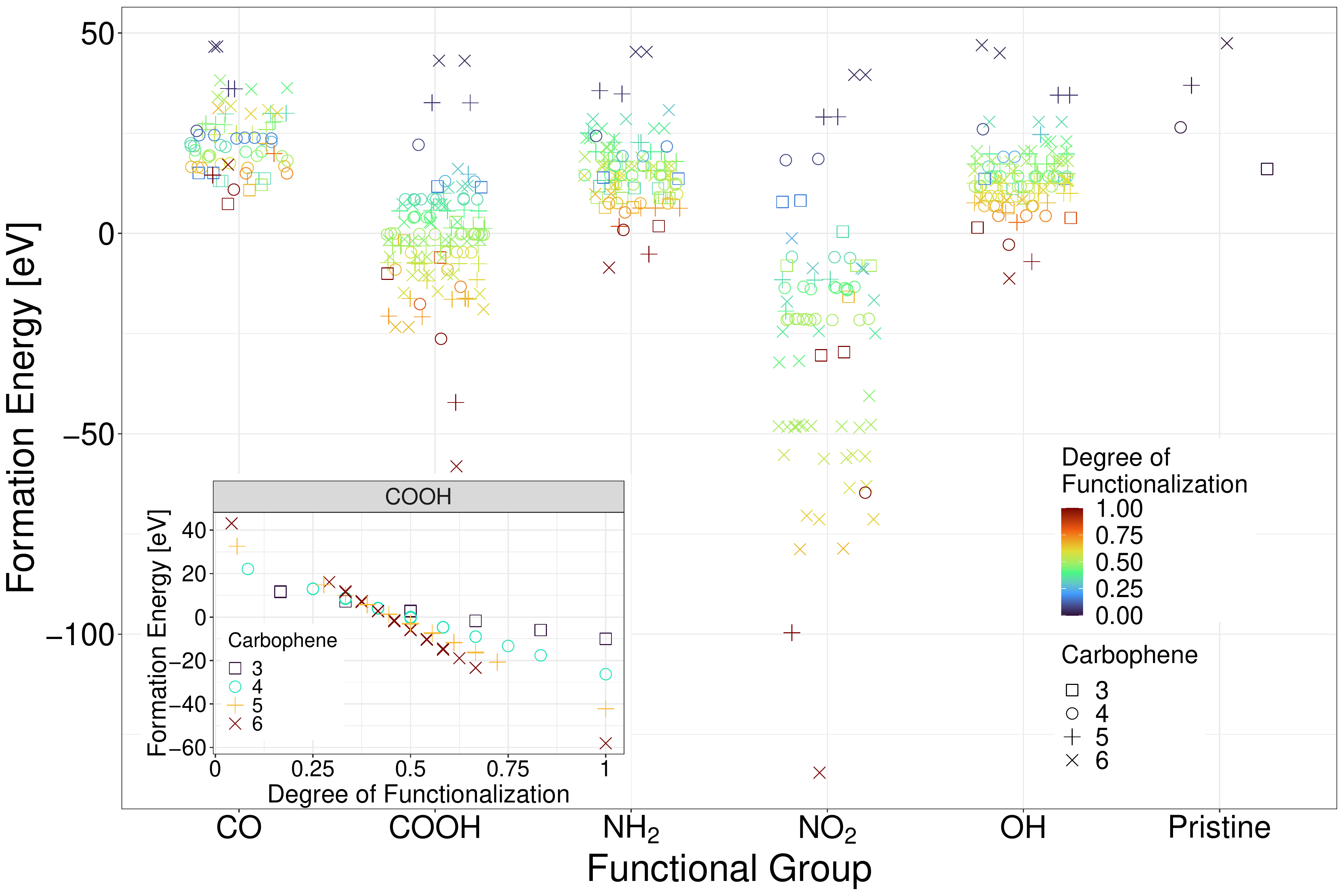}
\caption{Formation energies of N-carbophenes as the number of functional groups increase.  Insert: Formation energies of N-carbophenes as the number of COOH functional groups increase.}
\label{fig2:pfuncgroupvsFE}
\end{figure}

Figure \ref{fig2:pfuncgroupvsFE} displays how the formation energies of functionalized carbophenes decrease as more functional groups are added to a carbophene. 
Variations in the formation energies when two of the same functional groups are present in a cell are due to how they interact with each other or the nearby H atoms.  The strongest interactions are most likely due to Coulomb forces, but interactions mediated by the carbophene's $\pi$ electrons, as is found with adsorbates on graphene, may also be happening.\cite{Solenov13115502}  It is evident there is a substantial stabilizing effect for highly substituted carbophenes with NO$_2$ functionalities, with significantly lower formation energies than the rest. The stabilizing effect may be because the highly electron-withdrawing nitro groups stabilize the partial positive charges that result from neighboring nitro groups.
Using COOH as an example, the inset demonstrates the approximately linear dependence of formation energy on the degree of functionalization, indicating that differential formation energies per added group do not vary significantly with the degree of functionalization, which means that functionalization is likely to proceed (when successful) to a high degree of functionalization. Furthermore, this nearly linear trend happens with each studied functional group; see the Supplemental Information.
That the formation energies of carbophenes decrease with functional group adsorption, and in some cases have negative formation energies, could be another indication that Du \textit{et al.} synthesized a carbophene instead of graphenylene.  Their observation of oxygen in the XPS spectra could have been due to terminating functional groups.

Thus, for each functional group type, the first added functional reduces the formation energy by about the same amount for all carbophenes; notice that the degree of functionalization for the first group is different for each size of carbophene. For example, when all H atoms have been substituted, the formation energy of 6-carbophene has been reduced more than the other carbophenes. But the 6-carbophene has a higher H/C ratio; thus, more functionals are added per C atom than the 3-carbophene. Therefore, the greater stability gain at full functionalization is due to the 6-carbophene being able to accommodate more functional groups per C atom, which leads to greater energy reduction.

%%%%%%%%%%%%%%%%%%%%%%%%%%%%%%%%%%%%%%%%%%%%%%%%%%%%%%%%%%%%%%%%%%%%%%%%%%%%%%
%%%%%%%%%%%%%%%%%%%%%%%%%%%%%%%%%%%%%%%%%%%%%%%%%%%%%%%%%%%%%%%%%%%%%%%%%%%%%%
%%%%%%%%%%%%%%%%%%%%%%%%%%%%%%%%%%%%%%%%%%%%%%%%%%%%%%%%%%%%%%%%%%%%%%%%%%%%%%
%\subsection{Bands, PDOS, W}
%%%%%%%%%%%%%%%%%%%%%%%%%%%%%%%%%%%%%%%%%%%%%%%%%%%%%%%%%%%%%%%%%%%%%%%%%%%%%%
%%%%%%%%%%%%%%%%%%%%%%%%%%%%%%%%%%%%%%%%%%%%%%%%%%%%%%%%%%%%%%%%%%%%%%%%%%%%%%
%%%%%%%%%%%%%%%%%%%%%%%%%%%%%%%%%%%%%%%%%%%%%%%%%%%%%%%%%%%%%%%%%%%%%%%%%%%%%%

A previous work demonstrated that the valence and conduction bands of pristine carbophene are due to molecular orbitals centered on the p$_z$ atomic orbitals in pairs of 4-member ring carbon atoms.\cite{junkermeier2019simplecarbophene} The hydrogen s-orbitals play almost no role in the DOS around the Fermi level in pristine carbophene.  The carbon atoms to which the H are bound also play only a minor role in the DOS near the Fermi level.

\begin{figure}
\centering
\includegraphics[clip,width=5 in, keepaspectratio]{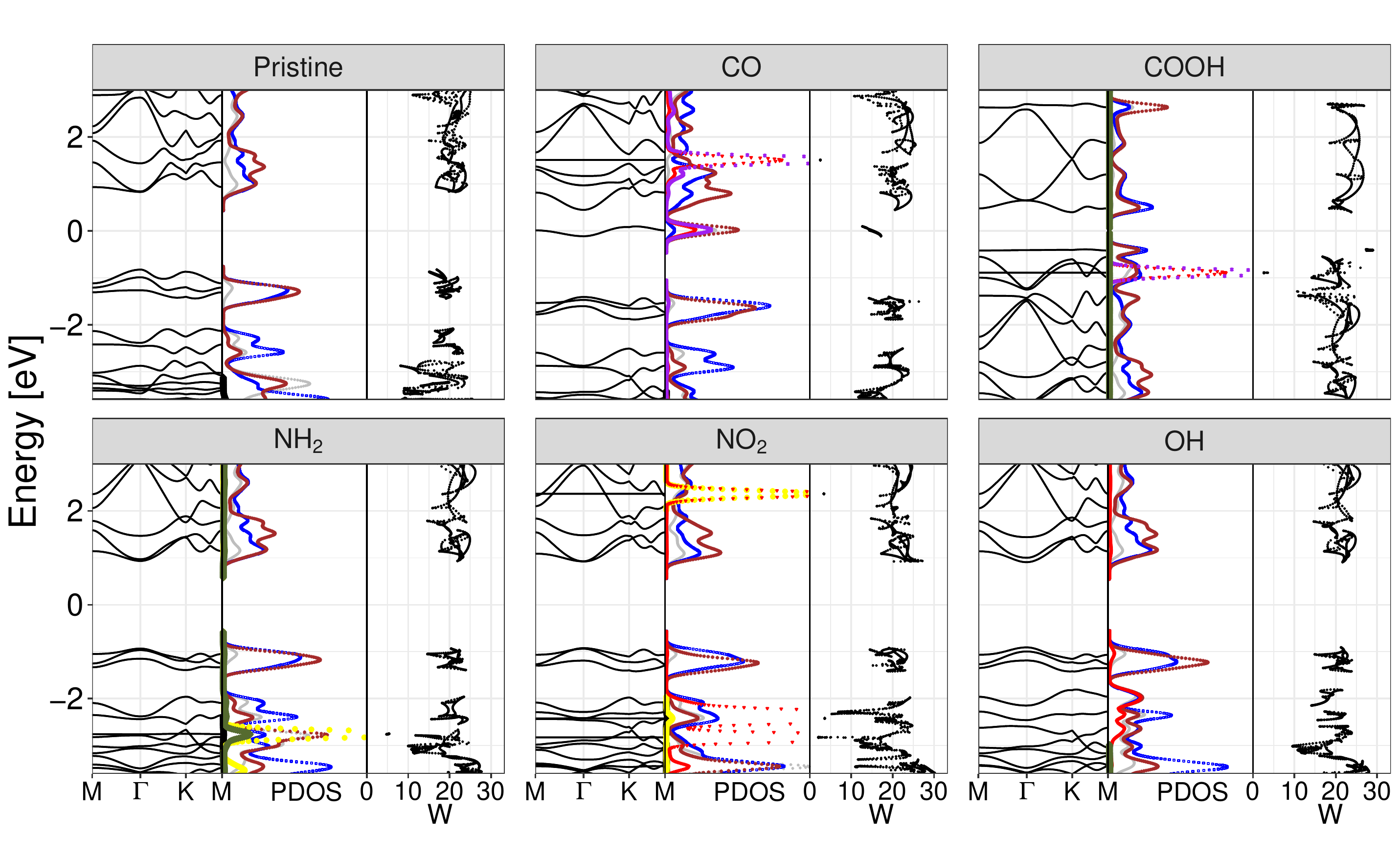}
\caption{Graphs containing band dispersion, PDOS, and the number of accessible atoms, W, of pristine 3-carbophene and 3-carbophenes with a single OH, NO$_2$, NH$_2$, or COOH functional group.  The colors and shapes used to represent specific types of atomic orbitals in the PDOS are specified in Fig. \ref{fig4:1pdoslegend}.}
\label{fig3:1functionalpdos}
\end{figure}

\begin{figure}
\centering
\includegraphics[clip,width=5 in, keepaspectratio]{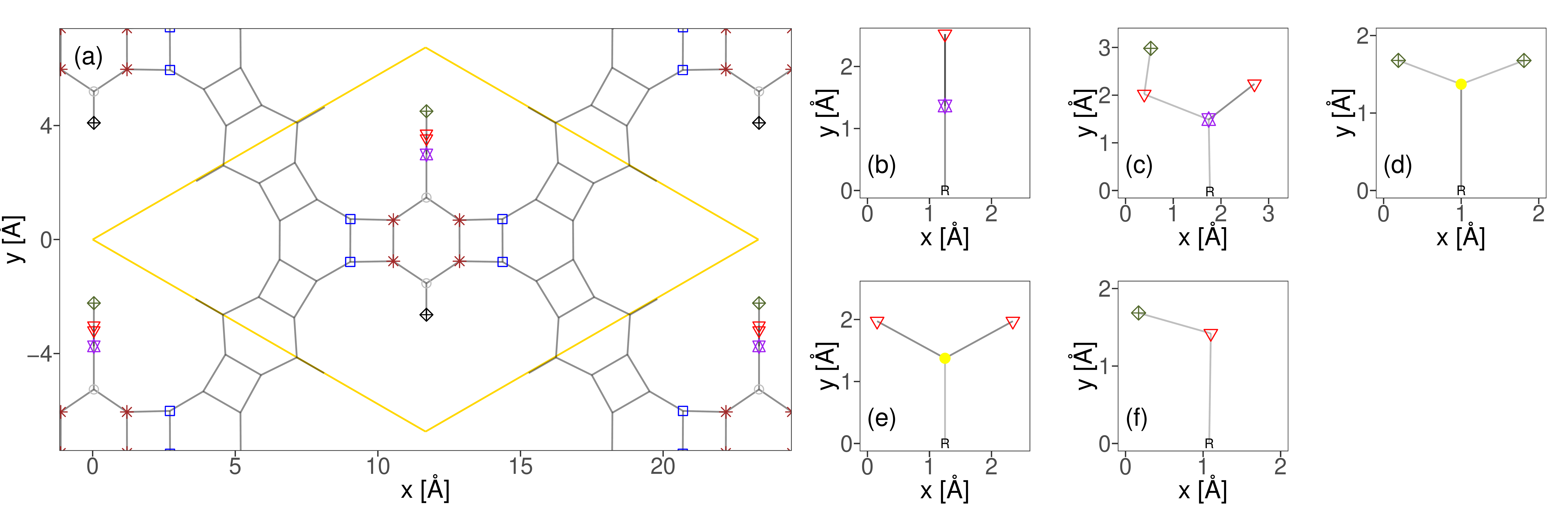}
\caption{PDOS legend for 3-carbophene with functional groups: (a) 3-carbophene with an NH$_2$ functional, (b) CO functional, (c) COOH functional, (d) NH$_2$ functional, (e) NO$_2$ functional, and (f) OH functional.  Except for H$_{s}$ orbitals, only the p orbitals are shown: C$_p$ in a functional:purple triangles, N$_p$:yellow filled-circles, O$_p$:red inverted triangles, H$_s$ in a functional:dark olive green diamond-plus, H$_s$ on carbophene:black diamond-plus, with blue squares, brown stars, and grey circles each being C$_p$ orbitals in different bonding structures.}
\label{fig4:1pdoslegend}
\end{figure}

In Figure \ref{fig3:1functionalpdos}, we compare the band dispersion, projected density of states (PDOS), and the number of accessible atoms, W, of pristine 3-carbophene with singly-functionalized 3-carbophenes.  Similar plots for functionalized 4-, 5-, and 6-carbophenes are given in the Supplemental Information.  As with graphene, the PDOS demonstrates that the bands near the Fermi level are primarily composed of the carbophene carbon p$_z$ orbitals.\cite{junkermeier2013highly}  The PDOS in Figure \ref{fig3:1functionalpdos} demonstrates that the atomic orbitals associated with most functionals play a small role in the region around the Fermi level. The outlier is the narrow, detached band at the Fermi level centered on the carbophene C atom, represented by the gray circle in Figure \ref{fig4:1pdoslegend}, bound to the C atom in the CO functional group.  The band pinned to the Fermi level may be due to charge transfer from the functional groups into the carbophene. A few functional groups pull charge out of the carbophenes (i.e., NH$_2$: 0.03 $e^{-}$, NO$_2$: 0.12 $e^{-}$, and OH: 0.07 $e^{-}$) which may widen the band gap.  COOH contributes only a small fraction of a charge to the carbophenes, 0.03 $e^{-}$, narrowing the gap. In contrast, CO functionals contribute 0.33 $e^{-}$ to a carbophene producing a mid-gap state pinned to the Fermi level.

The number of accessible atoms, W, is akin to the inverse participation ratio, and is a measure of the number of atoms on which an electron in state $\nu$ may be found.\cite{Junkermeier2008accessibleatoms}  Using the Boltzmann equation, we define
\begin{equation}
W(\nu) = e^{S(\nu)},
\end{equation}
where $S(\nu)$ is the entropy of information 
\begin{equation}
S(\nu) = - \sum_{i} p_{i}(\nu) \ln p_{i}(\nu),
\end{equation}
which is the sum over all atoms $i$ in the system, and $p_i(\nu)$ is the Mulliken population on atom $i$ in state $\nu$.  The range of $W$ is 1 to $N$, where $N$ is the number of atoms in the system. 
Practically, W allows the comparison of the distribution of elections in a way that is impractical using methods such as plotting electron density isosurfaces for many different states. 
Figure \ref{fig3:1functionalpdos}, includes the $W$ for each state at each kpoint. The figures demonstrates that states around the Fermi level are distributed over a large proportion of a unit cell's atoms. States that trap an electron into only a few functional group atoms occur deeper into the valence or conduction bands.

\begin{figure}
\centering
\includegraphics[clip,width=5 in, keepaspectratio]{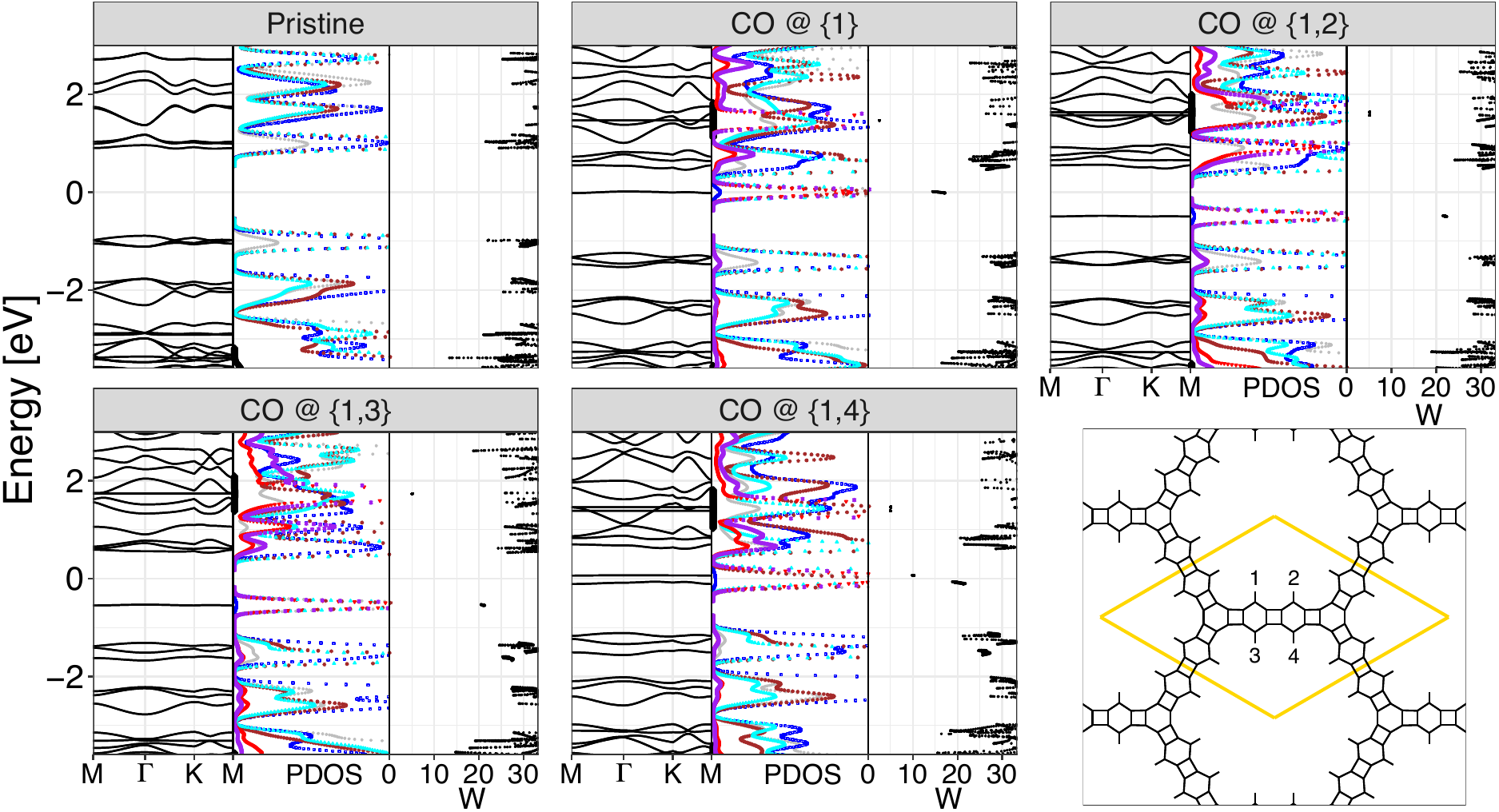}
\caption{Band structure, PDOS, and the number of accessible atoms of 4-carbophene in its pristine configuration, with one CO functional group, and with two CO functional groups.  The set of numbers in each graph title represent the bonding sites where the CO functionals were placed. The representation of a unit cell of pristine 4-carbophene gives the naming convention of the binding sites.}
\label{fig:4carbophene2CO}
\end{figure}

The interaction of CO functional groups with carbophenes illustrates how the other functional groups interact with the carbophene backbone.  Figure \ref{fig:4carbophene2CO} presents the band dispersion, PDOS, and W of 4-carbophene, in its pristine form and with one or two CO groups within the unit cell.  The bottom right frame of Figure \ref{fig:4carbophene2CO} gives the nomenclature we will use in describing where functional groups are attached (i.e., sites 1, 2, 3, or 4).  

A review of the bipartite graphene lattice and how it affects the bonding of multiple adsorbates on its surface is instructive in understanding how the interaction of CO functional groups with carbophenes changes the band structure.  
When a hydrogen (or fluorine) atom bonds to a particular atom in graphene, the charge transfer between that graphene atom and the H atom affects the graphene sublattices differently; neighboring carbon atoms of the same sublattice have partial charges of equal sign as the partial charge on the carbon to which the H atom is bound, while the neighboring atoms in the other sublattice have partial charges of the opposite sign.\cite{Junkermeierjp309419x}  The magnitude of the adsorbate induced partial charge decreases with n-th nearest neighbor order.  

Carbophenes do not have the sublattice nature of graphene, but a CO functional group at site 1 induces a partial charge in the C atom to which it is bound.  The nearest neighbors will, in turn, have induced charges of the opposite sign.  The next nearest neighbors will have the same sign, and so on.  If a second CO functional is placed at 4, the C atom to which it is bound already has a partial charge of the same sign as that induced.  Thus, these two groups produce nearly symmetric changes to the band structure, and we see a slight splitting of the bands at the Fermi level.  When the second functional group is at either bonding site 2 or site 3, the induced charge from the first CO group is now of the opposite sign to what the second CO functional wishes to induce, frustrating the band system and pushing the mid-gap states much further from the Fermi level.  

\begin{figure}
\centering
\includegraphics[clip,width=5 in, keepaspectratio]{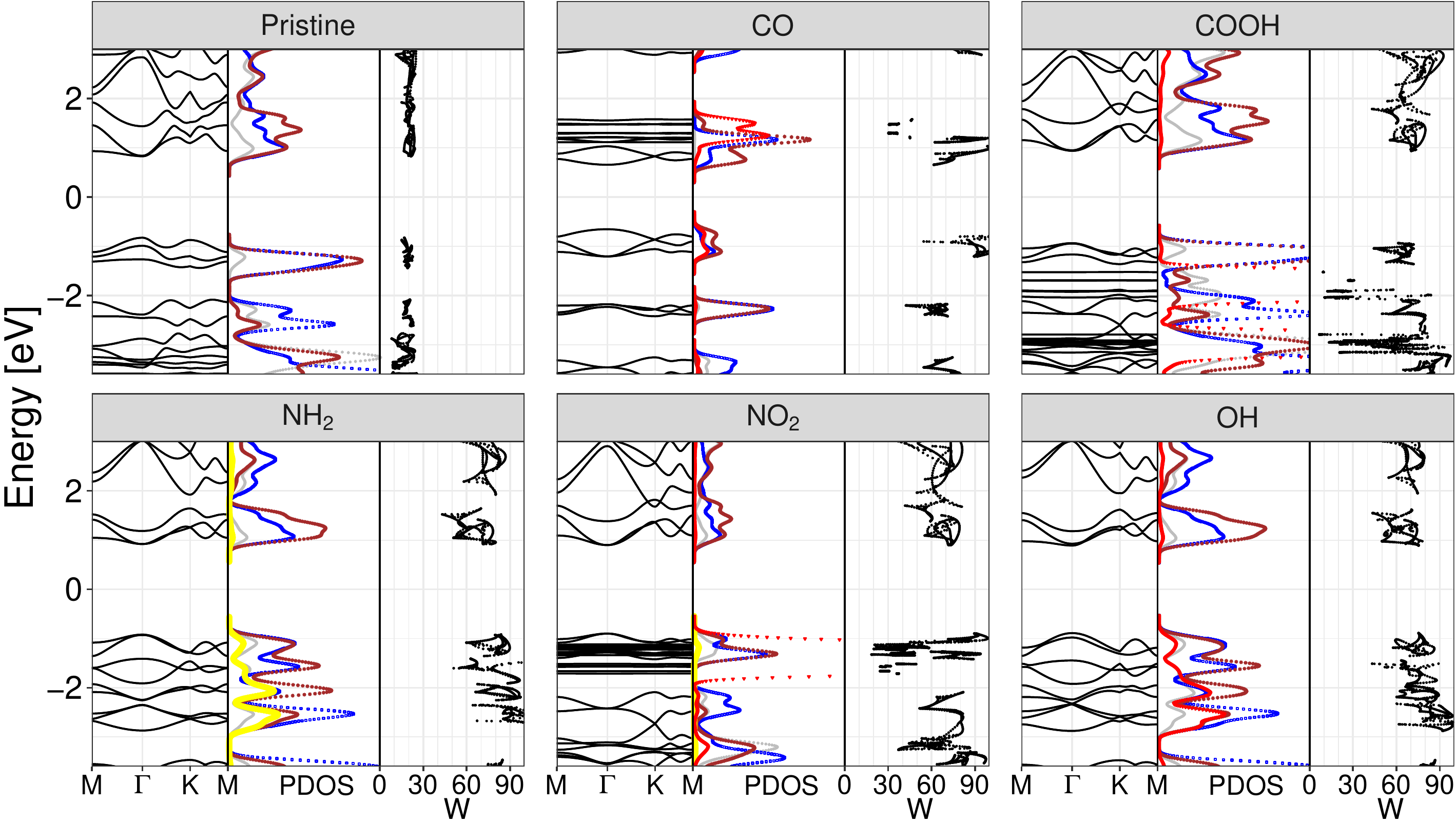}
\caption{Graphs containing band dispersion, PDOS, and the number of accessible atoms, W, of pristine 3-carbophene and 3-carbophenes with each H atom replaced by a functional group.  The colors and shapes used to represent specific types of atomic orbitals in the PDOS are specified in Fig. \ref{fig4:1pdoslegend}.}
\label{fig5:allfunctionalpdos}
\end{figure}

The band dispersion, PDOS, and W of fully functionalized 3-carbophenes are presented in Figure \ref{fig5:allfunctionalpdos}.  When all sites have a functional group, the groups occupy a significant part of the unit cell, providing more electrons to participate in the formation of energy levels. Thus, the functional states have a relevant effect on the overall distribution of orbitals, charges, and band structure. Therefore, the fully functionalized carbophenes all have band gaps. In addition, the functionals now have responses in the valence and conduction bands. The trends discussed for fully functionalized 3-carbophenes remain for fully functionalized 4-, 5-, and 6-carbophenes; see the Supplemental Information.

\begin{figure}
\centering
\includegraphics[clip,width=5 in, keepaspectratio]{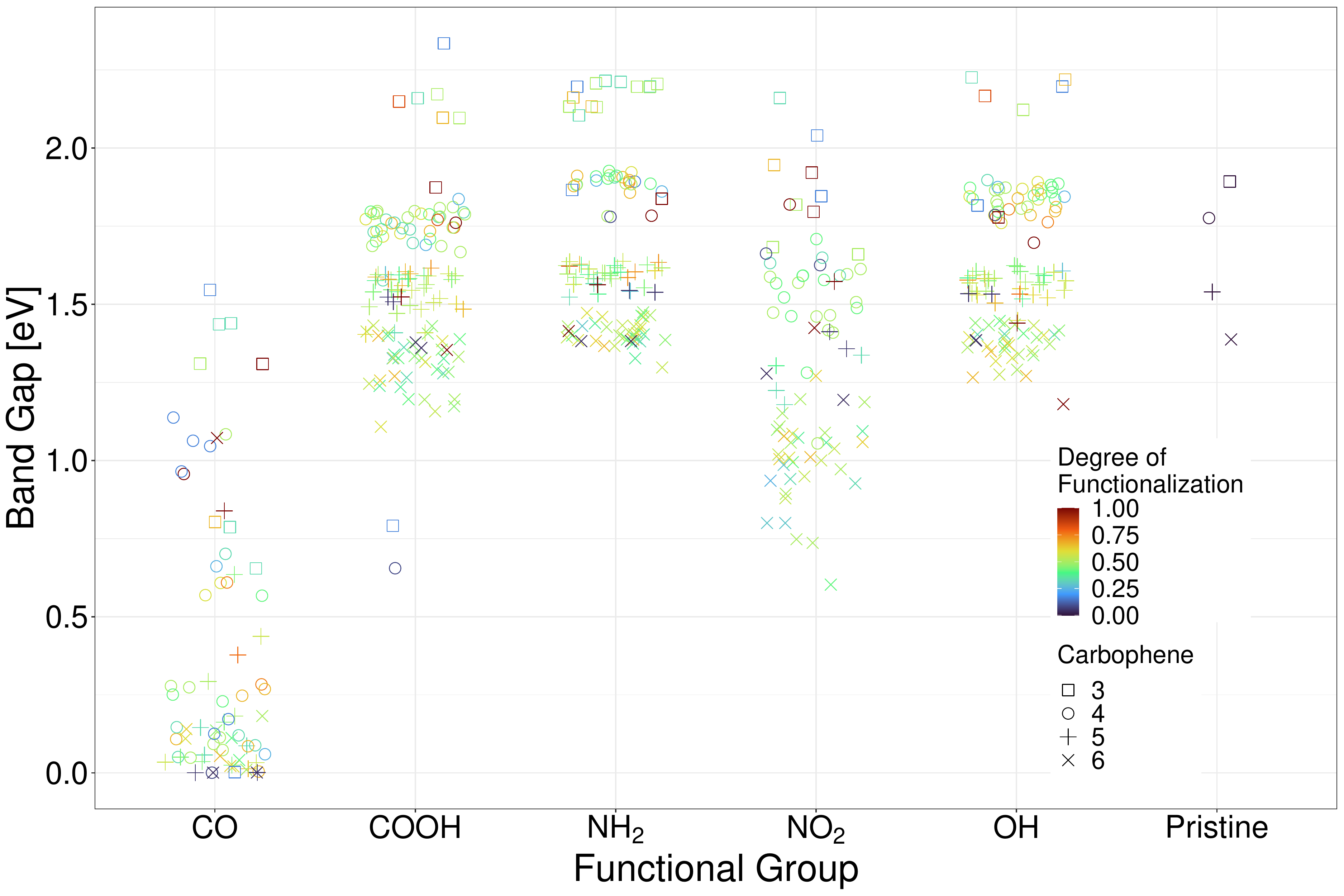}
\caption{Band gaps of N-carbophenes as the number of adsorbed functional groups increase.}
\label{fig:bandgaps}
\end{figure}

Figure \ref{fig:bandgaps} presents the computed band gaps of the randomly chosen binding site motifs that were discussed with Figure \ref{fig2:pfuncgroupvsFE}. As might be expected from the discussion of the two CO groups adsorbed on 4-carbophene, the bandgap changes as the number of functionals increase, but not necessarily in a linear fashion. Furthermore, except possibly in the case of the CO functional, the band gaps are stratified according to the functional type and type of carbophene.  Thus, it may be possible to tune a carbophene's band gap by choice of functionalization.

\section{Conclusion}

N-carbophenes are a possible outcome of a synthesis process designed to produce graphenylene.  Previous results demonstrated that carbophenes are energetically more favorable than graphenylene.  This work indicates that functionalization of carbophenes is energetically favorable, with the ability to functionalize each of the carbon dangling bonds. 
The presence of any of the considered functional groups is able to cause structural modifications on the carbophenes. Our results indicate CO group as the one which causes the smaller modification on bond lengths, followed by NH$_2$, NO$_2$, OH and finally COOH which brings the larger deformations on the structures.
Considering the effect of functional groups on formation energies, the increase of the degree of functionalization resulted in a linear decrease on formation energy, indicating an important stabilizing effect and suggesting that high occupations are energetically favorable.
Taking into account the effects on valence band and conduction band states, the functionalization altered the band gap opening for all carbophenes considered in the present study. Several states far from the frontier orbitals are affected by the presence of functional groups, but it is also possible to create band states close to the Fermi level, as is the case for CO functionalization for instance. Our results allow the observation of a large variation on band gap openings which is dependent on functional group and carbophene type considered. This last result can pave the way for a possible band gap tuning if the level of functionalization could be controlled experimentally.
At low levels of functionalization, most of the studied functional groups contributed bands away from the valence and conduction bands. However, the polar molecule carbonyl contributed a narrow band pinned to the Fermi level.  At complete functionalization, all functional groups contribute to the valence and conduction bands, but there are no longer any bands at the Fermi level.  These results suggest that carbophenes may be an excellent choice for applications where functionalized layered materials are needed.

\section*{Supplementary material}

Supplementary materials for this article are available online.

\section*{Acknowledgement}
Greenwood: A library for creating molecular models and processing molecular dynamics simulations was used in the production of the atomic structure models and data analysis.\cite{greenwood2018} Further data analysis was performed using R: A language and environment for statistical computing.\cite{RCoreBIB,ggplot2BIB,Hadleytidyverse,cowplotBIB} C. Junkermeier was supported by the United States funding agency NSF under Award Number 2113011, and by the University of Hawai`i Information Technology Services – Cyberinfrastructure. R. Paupitz acknowledges Brazilian funding agencies: FAPESP (grant \#2018/03961-5) and CNPq (grants \#437034/2018-6 and \#315008/2020-2) and computing resources supplied by the Center for Scientific Computing (NCC/GridUNESP) of the S\~ao Paulo State University (UNESP).

\section*{{CReDiT} authorship contribution statement}

Chad E. Junkermeier: Conceptualization, Data curation, Formal analysis, Investigation, Methodology, Project administration, Resources, Software, Validation, Visualization, Writing – original draft, Writing – review \& editing.
George Psofogiannakis: Writing – review \& editing.
Ricardo Paupitz: Resources, Writing – review \& editing.

% \appendix

%%%%%%%%%%%%%%%%%%%%%%%%%%%%%%%%%%%%%%%%%%%%%%%%%%%%%%%%%%%%%%%%%%%%%%%%%%%%%%
%%%%%%%%%%%%%%%%%%%%%%%%%%%%%%%%%%%%%%%%%%%%%%%%%%%%%%%%%%%%%%%%%%%%%%%%%%%%%%
%%%%%%%%%%%%%%%%%%%%%%%%%%%%%%%%%%%%%%%%%%%%%%%%%%%%%%%%%%%%%%%%%%%%%%%%%%%%%%
%%%%%%%%%%%%%%%%%%%%%%%%%%%%%%%%%%%%%%%%%%%%%%%%%%%%%%%%%%%%%%%%%%%%%%%%%%%%%%
%%%%%%%%%%%%%%%%%%%%%%%%%%%%%%%%%%%%%%%%%%%%%%%%%%%%%%%%%%%%%%%%%%%%%%%%%%%%%%
%%%%%%%%%%%%%%%%%%%%%%%%%%%%%%%%%%%%%%%%%%%%%%%%%%%%%%%%%%%%%%%%%%%%%%%%%%%%%%
\section*{Bibliography}

\bibliographystyle{iopart-num.bst}

\bibliography{carbophene}

\end{document}